\documentclass[12pt]{article}
\usepackage{amsmath,amssymb,amsthm,amsfonts}
\usepackage{amsmath,amssymb,makeidx,amsfonts,latexsym,enumerate}

\newcommand{\al}{\alpha}
\newcommand{\de}{\delta}

\title{Frequency derivation of the EPR-Bohm correlations}

\author{Andrei Khrennikov \\International Center for Mathematical
Modeling \\ in Physics and Cognitive Sciences,\\
University of V\"axj\"o, S-35195, Sweden\\
Email:Andrei.Khrennikov@msi.vxu.se}
\begin{document}
\maketitle
\begin{abstract}
The probabilistic structure of quantum mechanics is investigated in the frequency
framework.
By using rather complicated frequency calculations we reproduce the EPR-Bohm correlation
function which is typically derived by using the calculus of probabilities in a Hilbert
space. Our frequency probabilistic model of the EPR-Bohm experiment is a realist model --
physical observables are considered as objective properties of physical systems. It is also
local -- a measurement over one part of a composite system does not disturb another part
of this system. Nevertheless, our result does not contradict to the well known Bell's
``NO-GO'' theorem. J. Bell used the conventional (Kolmogorov) measure-theoretical approach.
We use the frequency approach. In the latter approach there are no reasons to assume that
the simultaneous probability distribution exists: corresponding frequencies may fluctuate
and not approach any definite limit (which Bell would like to use as the probability).
The frequency probabilistic derivation demonstrated that
incompatibility of observables under consideration plays the crucial role in
producing of the EPR-Bohm correlations.


\end{abstract}

\section{Introduction}
Since first years of quantum theory, unusual behaviour of probabilities in
experiments with quantum systems attracted attention of physicists, mathematicians
and even philosophers, see, e.g., [1]-[12] (see A. Holevo [11] for the recent
reanalyzing of the statistical structure of quantum theory). The central problem was a rather strange (unconventional)
behaviour of probabilities in the two slit experiment and other superposition-type
experiments.\footnote{According to Feynman et al. [13] this is ``... a phenomenon
which is impossible, absolutely impossible, to explain
in any classical way, and which has in it the heart of quantum mechanics. It really contains
the {\it only} mystery."} The conventional rule for addition of probabilities of
alternatives:
\begin{equation}
\label{F1}
P=P_1+P_2
\end{equation}
does not work in experiments with elementary particles. Instead of this rule, we have to use quantum rule:
\begin{equation}
\label{F2}
P=P_1+P_2+2\sqrt{P_1P_2}\cos\theta.
\end{equation}
This rule could be easily derived by using the method of Hilbert space. However,
there is a rather common viewpoint that this rule could not be obtained in the
conventional probabilistic framework. Typically it is said that quantum randomness
is {\it irreducible} -- in the opposite to classical randomness that can be
(at least in principle) reduced to randomness of initial conditions and
perturbations, see e.g. A. Zeilinger [14]  for an extended discussion.\footnote{However,
compare with, e.g., A. Holevo [11].}

Such a viewpoint to quantum randomness was strongly supported by investigations
on the EPR-Bohm correlations. The crucial step was done by J. Bell [15] who by
proving his inequality demonstrated that quantum correlations could not be reduced
to (local) classical correlations.

In [12] I performed the careful analysis of standard considerations on quantum probabilities.
This analysis demonstrated that the main source of many quantum misunderstandings
is vague manipulation with quantum probabilities. Typically physicists (including J. Bell
and many others investigating the EPR-Bohm experiment) as well as
mathematicians operate with the symbol ${\bf P}$ of an abstract probability measure.
This symbol has no direct relation to a concrete experimental situation. However,
already N. Bohr pointed out that in quantum theory the whole experimental arrangement
should be taken into account. Unfortunately N. Bohr was concentrated merely on dependence
of {\bf individual} quantum events on experimental conditions. In particular, the fundamental
notion of {\it Bohr's experimentalism} is the notion of {\it phenomenon,} [16], [17]. Here a phenomenon
is an individual event which is determined by the interaction of a quantum system with
a measurement apparatus. He discussed the two slit experiment. Here a dot on the registration screen
when both slits are open is one phenomenon. A dot when just one slit is open is another
phenomenon.

Of course, the introduction of the notion of phenomenon was of the greatest importance.
However, quantum theory does not provide any description of individual events. This is
a {\it statistical theory.} Therefore it was essentially more important to underline from the
very beginning that quantum probabilities (and not only the results of individual measurements)
depend on complexes of experimental physical conditions. Unfortunately it was not done
neither by N. Bohr nor by any of his successors. In particular, this induced a rather mystical
viewpoint to quantum probabilities as totally different from conventional (classical) probabilities.

Starting with the (more or less) evident fact that in general probabilities should
depend on complexes of experimental physical conditions -- {\bf contexts} -- I developed
[18] a {\it contextual approach to quantum probabilities.} It was demonstrated
that the quantum interference rule (\ref{F2}) can be easily derived in the contextual probabilistic
framework.\footnote{My investigations were not the first contextual investigations on quantum probabilities,
see,  e.g., L. Accardi [19] -- camelion effect (and the corresponding applications to the EPR-experiment), or
S. Gudder [20] -- the theory of probability manifolds.}

In the present paper I present a contextual (frequency) probabilistic derivation of expressions coinciding with
the EPR-Bohm
correlation functions. It is demonstrated that (in the opposite to a rather common opinion)
those correlations can be obtained in the {\it local realist} (but contextual) approach
if we carefully combine probabilities corresponding to different physical contexts.

Our contextual derivation of the EPR-Bohm covariations does not contradict to Bell's arguments and
their generalizations, see  e.g. [21]. The original Bell arguments were based  on
calculations with an abstract (context independent) probability distribution ${\bf P}$
on the space of hidden variables. Such calculations
are impossible in our contextual probabilistic framework. Another way to obtain Bell's type
inequalities is to use {\bf counterfactuals}, see, e.g., [21]. This way is also closed in the
contextual probabilistic framework.

As in our previous papers [18], we use the {\it frequency}
contextual probabilistic framework. Probabilities are defined
as limits of frequencies in long runs of experiments. Such frequency
probabilities directly depend on experimental physical conditions.

Our frequency framework is {\bf contextual} and it is fundamentally different
from the frequency framework which was used by many authors, see, e.g. Stapp and Eberhard
[21], to derive Bell's inequality. Stapp-like framework is noncontextual. In particular,
people freely operated with counterfactual statistical data. The critical analysis of
the use of counterfactual data in the EPR-Bohm model was performed by W. De Baere [22] who demonstrated
that there was no physical justification of the use of counterfactuals.

\section{Contextual frequency viewpoint to statistical measurements over composite
systems.}
Let us consider a preparation procedure ${\cal E}$ that produces a statistical ensemble $S$
of physical (or biological, or social) systems, $\omega \in S.$
We suppose that each element $\omega \in S$ has two properties $a$ and $b$
represented by dichotomous variables $a(\omega)=a_1$ or $a_2$ and $b(\omega)=b_1$ or $b_2.$
We suppose that each of properties $a$ and $b$ is observable: values $a(\omega)$ and $b(\omega)$
can be measured by some measurement procedures ${\cal M}_a$ and ${\cal M}_b,$ respectively.

{\bf Remark 2.1.} (Realism) We use the {\bf realists approach to physical observables.} They are
considered as properties of an object (physical system). In the mathematical framework this
means that values of physical observables can be represented as functions
\begin{equation}
\label{FFF}
a=a(\omega), b=b(\omega),....
\end{equation}
of a parameter $\omega$  describing a physical system. Such a mathematical model
can be called {\it functional realists model.}
From the very beginning it is important
to underline the fundamental difference between this mathematical model  and
the model which was proposed by J. Bell (and widely used in the EPR-Bohm framework).
Bell's model can be called {\it measure-theoretical realists model.} In Bell's
model it is also supposed that physical observables can be represented in the functional form
(\ref{FFF}). But it is not the end of the story. The second Bell's fundamental assumption is that
there exists a probability measure ${\bf P}$ on the space of parameters $\omega$ such that all physical
observables can be mathematically represented as {\it random variables} on one fixed (Kolmogorov)
probability space. So Bell's realism is essentially stronger than our functional realism.
We shall see that in our ``weak-realists" model we can reproduce quantum mechanical expressions
for the EPR-Bohm correlations. By Bell's theorem it is impossible in his ``strong-realists" model.

In general we could not perform the measurement of the $a$-observable without to disturb
the system $\omega.$ Mathematically such a disturbance can be described by some
transformation $\omega \to \tilde \omega$ of a probability space. In general, for another observable $b,$
the probability distribution
of $b(\omega)$ could differ from the probability distribution of $b(\tilde \omega).$ \footnote{Thus we continue
by using Heisenberg's  viewpoint that mutual perturbations should be taken into account in
theory of measurement. W. Heisenberg discussed this problem in the context of quantum measurements.
However, there are no reasons that such disturbance effects could be important only in experiments with
quantum systems.}
The same is valid for the $b$-measurement: it also disturbs the system $\omega, \omega \to \hat{\omega}.$
In general the $a$ and $b$ properties cannot be measured simultaneously. We cannot create such a
measurement device ${\cal M}_{ab}$ that will give us the pair $(a(\omega), b(\omega))$
for the fixed element (e.g. a physical system) $\omega\in S.$

Let us now consider two preparation procedures, ${\cal E}$ and ${\cal E}^\prime.$
They produce statistical ensembles, $S$ and $S^\prime$ of physical (or biological, or social)
systems, $\omega \in S$ and $\omega^\prime \in S^\prime.$
Elements of $S$ and $S^\prime$ have properties $a(\omega), b(\omega)$ and
$a^\prime(\omega^\prime), b^\prime(\omega^\prime),$ respectively.
For elements $\omega$ (respectively, $\omega^\prime),$
there are well defined two dichotomous observables $a=a_1, a_2$ and $b=b_1, b_2$
(respectively, $a^\prime=a_1^\prime, a_2^\prime$ and $b^\prime=b_1^\prime, b_2^\prime).$
However, in general $a(\omega)$ and  $b(\omega)$ (or $a^\prime(\omega^\prime)$ and
$b^\prime(\omega^\prime))$ cannot be measured simultaneously for fixed $\omega \in S$
(or $\omega^\prime \in S^\prime),$ see above considerations.

We shall use following statistical ensembles:

Ensembles $S_a(k), S_{a^\prime}^\prime(l)$ which are obtained from the
ensembles $S$ and $S^\prime,$ respectively, by using selective
procedures (filters) with respect to values $a=a_k$ and $a^\prime=a_l^\prime,$
respectively. We remark that the probability distributions of $b$ and $b^\prime$
for measurements performed over elements of
ensembles $S_a(k)$ and  $S_{a^\prime}^\prime(l)$ can essentially differ
from  distributions  for corresponding sub-ensembles of ensembles
$S$ and $S^\prime:$
$$
S_{0;a}(k)= \{ \omega \in S: a(\omega)=a_k\}, \;\mbox{and}\;
S_{0;a}^\prime(l)=\{ \omega^\prime \in S^\prime: a^\prime(\omega^\prime)=a_k^\prime\}.
$$
According to  W. Heisenberg in general the selections $a=a_k$ and $a^\prime=a_l^\prime$
can essentially change probability distributions of  other observables
(e.g., $b$ and $b^\prime).$ Probability distributions of the $b$ and $b^\prime$ for the ``hidden
sub-ensembles" $S_{0;a}(k),S_{0;a}^{\prime}(l)$ can essentially differ  from
probability distributions for the selected ensembles $S_a(k), S_a^{\prime}(l).$

{\bf Remark 2.2.} (Bohr's complementarity and Heisenberg's uncertainty) It is well
known that N. Bohr proposed the principle of complementarity on the basis of intensive
discussions with W. Heisenberg, see, e.g., [2]. The derivation of uncertainty relations by
W. Heisenberg was of the great importance for N. Bohr. It may be less known that
(at least from the beginning) views of N. Bohr and W. Heisenberg were essentially different .
Heisenberg's uncertainty principle says that a measurement of,  e.g., the position
$q$ causes  an uncontrollable disturbance of the momentum $p$
and vice versa. Bohr's complementarity principle says that it is totally meaningless
even consider the momentum $p$ in the experimental arrangement for a
$q$-measurement. From Bohr's viewpoint, in this paper we are doing totally forbidden
things. However, from Heisenberg's viewpoint, our considerations look quite natural.

For a finite set $O,$ the number of elements in $O$ is denoted
by the symbol $\vert O \vert.$ We set
$$
n_a(k) =\vert S_a(k) \vert, \;  n_{a^\prime}(l) =\vert S_{a^\prime}^\prime(l) \vert\;.
$$
We shall also use numbers:
$$
 n_{b/a}(i/k)\equiv n_b(i; S_a(k))=\vert \{ \omega \in S_a(k):b(\omega)=b_i\}\vert,
 $$
 $$
n_{b^\prime/a^\prime}(j/l)\equiv
n_{b^\prime}(j; S_{a^\prime}^\prime(l))=\vert \{ \omega^\prime \in S_{a^\prime}^\prime(l)):
b^\prime(\omega^\prime)=b_j^\prime\}\vert\;,
$$
that are numbers of elements in the ensembles
$S_a(k)$ and $S_{a^\prime}^\prime(l),$ respectively,
for that $b=b_i$ and $b^\prime=b_j^\prime$, respectively.
We now introduce following relative frequencies with respect to different ensembles:
$$
\nu_{b/a}(i/k)\equiv \nu_b(i; S_a(k))=
\frac{n_b(i; S_a(k))}{n_a(k)},
$$
the frequency to get $b=b_i$
in the ensemble $S_a(k).$
This frequency can be called the {\bf conditional frequency} of $b=b_i$ under the condition
$a=a_k.$ However, we prefer to call it the {\bf contextual frequency} to distinguish  conditioning
with respect to a context (given by a new ensemble $S_a(k))$ and conventional conditioning
(used, e.g., in the Kolmogorov measure-theoretical model) based on Bayes' formula for the conditional
probability. The conventional approach to conditioning is not contextual. In the conventional
approach we should fix from the beginning one single Kolmogorov probability measure ${\bf P}$ and then operate
with conditional probabilities with respect to this fixed measure. In the conventional approach:
$$
{\bf P}(b=b_i/a=a_k)= {\bf P}(b=b_i,a=a_k)/ {\bf P}(a=a_k).
$$

In the same way we define the contextual frequency for $a^\prime$ and
$b^\prime,$
$
\nu_{b^\prime/a^\prime}(j/l) \equiv  \nu_{b^\prime}(j; S_{a^\prime}^\prime(l)).
$

In the case when we should underline ensemble dependence (i.e., context dependence)
of frequencies we
will use the symbols $\nu_b(i; S_a(k)), \nu_{b^\prime}(j; S_{a^\prime}^\prime(l))$
and so on. In technical calculations we will omit ensemble dependence and use
symbols $\nu_{b/a}(i/k), \nu_{b^\prime/a^\prime}(j/l)$ and so on.

Suppose that there exists a preparation procedure $G$ which produces pairs
$w=(\omega, \omega^\prime)$ of systems, {\it composite systems,} such that for each fixed
$\omega^\prime$ observations over
$\omega$ produce the same statistics as observations over $\omega \in S$ and vice versa.
The $G$ produces a statistical ensemble ${\cal S}$ of pairs $w=(\omega, \omega^\prime).$ \footnote{So
the restriction to the preparation procedure $G$ is that marginal distributions
with respect to $\omega$ and $\omega^\prime$ systems coincide with distributions with respect
to ensembles $S$ and $S^\prime$ produced by preparation procedures
${\cal E}$ and ${\cal E}^\prime.$ }

In particular, we can choose as a $G$ some preparation procedure which is used
for {\bf preparation of the EPR-pairs in the EPR-Bohm experiment.} However, at the moment
we consider a more general framework.

We consider following properties of elements of ${\cal S}:{\bf a}(w)=
(a(\omega), a^\prime (\omega^\prime))$ and ${\bf b}(w) =(b(\omega), b^\prime(\omega^\prime)).$
We shall use following sub-ensembles of ${\cal S}:$
$$
{\cal S}_{0;{\bf a}} (kl)=
\{ w=(\omega,\omega^\prime)  \in {\cal S}: a(\omega)= a_k, a^\prime(\omega^\prime)= a_l^\prime\}\;.
$$
We suppose that properties ${\bf a}=(a, a^\prime)$ and ${\bf b}=(b, b^\prime)$ are observable:
for any $w=(\omega, \omega^\prime) \in {\cal S}, $ we can measure both $a(\omega)$ and $a^\prime(\omega^\prime)$
(or $b(\omega)$ and $b^\prime(\omega^\prime)).$ Thus a measurement over the part $\omega$ of the system $w$
does not disturb the part $\omega^\prime$ of the system $w$ and vice versa.
In particular, such a situation we have in the EPR experiment for correlated quantum particles.
In the EPR experiment
we can escape mutual disturbances by using  {\bf spatial separation of the parts}
$\omega$ and $\omega^\prime$ of
the composite system $w=(\omega, \omega^\prime).$ \footnote{However, spatial separation
is only the sufficient condition under that the ${\bf a}$ and ${\bf b}$ are observable. In general
spatial separation need not be involved in our considerations (at least for macroscopic systems).}

We shall also use the statistical ensembles ${\cal S}_{\bf a}(kl)$ that are obtained from the ensemble ${\cal S}$ by using
selective procedures (filters) with respect to values ${\bf a}=(a=a_k, a^\prime= a_l^\prime).$
We remark that the distributions of ${\bf b}$ for elements of the ensemble ${\cal S}_{0;{\bf a}} (kl)$
and the ensemble ${\cal S}_{\bf a}(kl)$ can differ essentially. The preparation of the later ensemble
disturbs composite systems. Nevertheless, we can assume that (at least for large ensembles)
\begin{equation}
\label{EQR}
n_{\bf a}^0(kl) \equiv \vert {\cal S}_{0;{\bf a}} (kl) \vert=
\vert  {\cal S}_{\bf a}(kl) \vert \equiv n_{\bf a}(kl),
\end{equation}
since we create the ensemble ${\cal S}_{\bf a}(kl)$ by selecting from the ensemble ${\cal S}$
elements belonging to the ensemble ${\cal S}_{0;{\bf a}}(kl).$ In our present model
the only disturbing feature
of this procedure is the change of the ${\bf b}$-distribution.\footnote{In principle,
we could study more general models in that $n_{\bf a}^0(kl) \not= n_{\bf a}(kl)$
(even approximately).}

In the same way we introduce ensembles ${\cal S}_{\bf b}(ij)$ and numbers
$n_{\bf b}(ij).$
Finally, we consider
$$
n_{{\bf b}/{\bf a}}^0(ij/kl),\; \mbox{and} \; n_{{\bf b}/{\bf a}}(ij/kl)
$$
numbers of elements in the ensemble   ${\cal S}_{0;{\bf a}}(kl)$ and the ensemble
 ${\cal S}_{\bf a}(kl)$, respectively,
for which ${\bf b}=(b_i, b_j^\prime).$

We now introduce following relative frequencies with respect to different ensembles:
$$
\nu_{\bf a}(kl)
=\frac{n_{\bf a}(kl)}{M}, M=|{\cal S}|,
$$
-- the frequency to get ${\bf a} =(a_k, a_l^\prime)$
in the ensemble ${\cal S}$ and
$$
\nu_{{\bf b}/{\bf a}}^0(ij/kl)=
\frac{n_{{\bf b}/{\bf a}}^0(ij/kl)}{n_{\bf a}(kl)},\; \mbox{and}\;
\nu_{{\bf b}/{\bf a}}(ij/kl)=
\frac{n_{{\bf b}/{\bf a}}(ij/kl)}{n_{\bf a}(kl)},
$$
-- the frequencies to get ${\bf b}=(b_i, b_j^\prime)$ in the ensemble ${\cal S}_{0;{\bf a}}(kl)$
and the ensemble ${\cal S}_{\bf a}(kl),$ respectively. These are contextual frequencies --
to observe the ${\bf b}=(b_i, b_j^\prime)$ in the contexts ${\cal S}_{0;{\bf a}}(kl)$ and
${\cal S}_{\bf a}(kl),$ respectively.

We remark that $\nu_{{\bf b}/{\bf a}}^0(ij/kl)$
are ``hidden frequencies". We could find them only if it was possible to eliminate the
perturbation effect of the ${\bf a}$-selection. This can be done for classical systems,
i.e., physical systems which are not sensitive to perturbations corresponding to selections.
However, in the general case (in particular, for quantum systems) we cannot eliminate
effects of perturbations.

We also use frequencies:
$\nu_{\bf b}(ij)=\frac{n_{\bf b}(ij)}{M},$ the frequency
to get ${\bf b}=(b_i, b_j^\prime)$ in the original ensemble ${\cal S}.$
Finally, we consider frequencies
$$
\nu_{\bf ba}(ijkl) =\frac{n_{\bf ba}(ijkl)}{M},
$$
where $n_{\bf ba}(ijkl)$ is the number of elements in the ensemble
${\cal S}$  for that
$$
b=b_i, b^\prime=b_j^\prime, a=a_k, a^\prime=a_l^\prime.
$$
We notice that the quadruple
$({\bf b, a})=(b, b^\prime, a, a^\prime)$ need not be an observable, compare to [22].
For example, we could
not observe $(\bf b, a)$ if
$(b, a)$ or $(b^\prime, a^\prime)$ are not observable.
Thus frequencies $\nu_{\bf ba}(ijkl; {\cal S})$ are not observable (they are ``hidden").

Since $a, b, {\bf a, b}$ are observables, we can use the principle of
{\it statistical stabilization for corresponding frequencies.} These frequencies
should converge to corresponding probabilities (when we repeat preparation and measurement
procedures many times):
$$
p_{b/a}(i/k)=\lim \nu_{b/a}(i/k)
$$
and analogous for $b^\prime;$ and also:
$$
p_{{\bf a}}(kl)
=\lim \nu_{\bf a}(kl);\;
p_{{\bf b}}(ij)
=\lim \nu_{\bf b}(ij).
$$
$$
p_{{\bf b/a}}(ij/kl)
=\lim \nu_{{\bf b}/{\bf a}} (ij/kl).
$$
It should be noticed that in general we can
not assume that frequencies $\nu_{{\bf ba}}(ijkl)$ stabilize!
So probabilities ${\bf P}(b=b_i, b^\prime=b_j^\prime, a=a_k, a^\prime=a_l^\prime)$
may be not exist at all!
In my former  probabilistic investigations on foundations of quantum mechanics [12] there were modeled
situations when the absence of the simultaneous probability distribution (chaotic fluctuations
of corresponding frequencies) did not contradict to the existence of probability distribution
(i.e., stabilization of frequencies to some limits) for observable quantities.

As we have already mentioned, to underline the ensemble dependence we will often use
symbols
$$
\nu_{\bf b}(ij; {\cal S}) [\equiv \nu_{{\bf b}}(ij)], \; \; \nu_{\bf a}(kl; {\cal S})
[\equiv \nu_{{\bf a}}(kl)]
$$
$$
\nu_{\bf b}(ij; {\cal S}_{0;{\bf a}}(kl)) [\equiv \nu_{\bf b}^0(ij;kl)],\;\;
\nu_{\bf b}(ij; {\cal S}_{\bf a}(kl))) [\equiv \nu_{\bf b}(ij;kl)]
$$
and so on. We understood that the reader is already tired by considering a large number of various frequencies.
This is one of disadvantages of the frequency approach (see R. von Mises [23] and my book [12] for detail).
However, the detailed frequency analysis is
the only possible way to provide correct understanding of the experimental situation.

By using Bayes-framework (see von Mises for corresponding frequency
considerations [23]) we can represent frequencies
for ${\bf b}=(b_i, b^\prime_j)$ in the ensemble
${\cal S}$ in the following way
$$
\nu_{{\bf b}}(ij)=\frac{n_{{\bf b}}(ij)}{M}=
\frac{1}{M}\sum_{k,l=1}^2 n_{{\bf  b a}}(ijkl)=
\sum_{k,l=1}^2 \nu_{{\bf ba}}(ijkl)
$$
$$
=\sum_{k,l=1}^2 \nu_{{\bf a}}(kl) \nu_{{\bf b}/{\bf a}}^0 (ij/kl).
$$
However, in general we could not proceed in classical-like way, namely
to take the limits of all frequencies on both sides of this equality.

Here the frequencies
$\nu_{\bf b}(ij; {\cal S}), \nu_{\bf a}(kl; {\cal S})$ have limits (since quantities
${\bf b}$ and ${\bf a}$ are observable), but the frequencies
$\nu_{\bf b} (ij; {\cal S}_{0;{\bf a}}(kl))$ need not. This is a consequence of
the fact that the frequencies $\nu_{\bf ba}(ijkl; {\cal S})$ need not stabilize.
But even if they stabilize the corresponding probabilities are not observable.
Therefore such probabilities
should be eliminated from considerations.

On the other hand, we know that the frequencies $\nu_{\bf b}(ij; {\cal S}_{\bf a}(kl))$
definitely stabilize. So we can modify the Bayesian framework by using latter frequencies.
Taking into account the ensemble dependence, we write:
\[\nu_{\bf b}(ij; {\cal S})=\sum_{k, l=1}^2 \nu_{\bf a} (kl; {\cal S})
\nu_{\bf b}(ij; {\cal S_{\bf a}}(kl))+ \delta(ij; {\cal S, {\cal S}_{\bf a}}),\]
where
\[\de (ij; {\cal S, \cal S}_{\bf a})
=\sum_{k,l=1}^2 \nu_{\bf a}(kl; {\cal S})
[\nu_{\bf b}(ij, {\cal S}_{0;{\bf a}}(kl))-\nu_{\bf b}(ij; {\cal S}_{\bf a}(kl))]\]
is an {\it disturbance term} which is induced by the transition from the ensemble
${\cal S}$ to selected ensembles ${\cal S}_{\bf a}(kl).$ We remark that $\delta=\delta^{(M)},$
where $M$ is the number of particles in the ensemble ${\cal S}.$
 By taking the limit when $M \to \infty$ we get:
\begin{equation}
\label{TRD}
p_{\bf b}(ij)=\sum_{k,l=1}^2 p_{\bf a}(kl)
p_{\bf b/a}(ij/kl)+\bar{\delta}_{{\bf b/a}}(ij),
\end{equation}
where
\begin{equation}
\label{TRDL}
\bar{\delta}_{\bf b/a}(ij) = \lim_{M\to \infty} \delta^{(M)}.
\end{equation}

Probabilities $p_{\bf b/a}(ij/kl)= \lim_{M\to \infty} \nu_{\bf b}(ij; {\cal S}_{\bf a}(kl))$
can be considered as frequency conditional probabilities. However, we prefer to call them
{\it contextual probabilities,}  see the previous discussion on relative frequencies.

{\bf Remark 2.3.} (Bell-Kolmogorov realism) By using Bell's approach
to realism we would obtain
the conventional formula of total probability:
\begin{equation}
\label{TRDT}
p_{\bf b}(ij)=\sum_{k,l=1}^2 p_{\bf a}(kl)
p_{\bf b/a}(ij/kl).
\end{equation}
It would be the end of the story: we would not be able to proceed and to obtain
the EPR-Bohm type correlation functions.

{\bf Remark 2.4.} (Entanglement) The presence of a nontrivial disturbance term
$\bar{\delta}_{\bf b/a}(ij)\not =0$ for composite systems is related to the phenomenon which
is known in the conventional quantum formalism as entanglement. We can say that if
$\bar{\delta}_{\bf b/a}(ij)\not =0$ for a composite system $w=(\omega,\omega^\prime)$
then the parts $\omega$ and $\omega^\prime$ of such a system are entangled. If $\bar{\delta}_{\bf b/a}(ij)=0$
then they are disentangled.

We now formulate the above result -- representation (\ref{TRD}) -- as the mathematical proposition.
Let $\{ {\cal S}^{(M)}\}$ be a sequence of ensembles such that
$\vert {\cal S}^{(M)}\vert =M.$ Let
$\{ {\cal S}_{\bf a}^{(M)}\}$ be a family (depending on the parameter ${\bf a}
=(a_k, a_l^\prime))$ of sequences of ensembles such that
$a(\omega)=a_k$ and $a^\prime(\omega^\prime)=a_l^\prime$ for any
$w=(\omega, \omega^\prime) \in {\cal S}_{\bf a}^{(M)}$ and the equality
(\ref{EQR}) holds true. In general ensembles ${\cal S}_{\bf a}^{(M)}$ have no
special relation to ensembles ${\cal S}^{(M)}.$  The equation (\ref{EQR})
connecting numbers of elements in the corresponding ensembles is the unique
constraint coupling those ensembles.

\medskip

{\bf Proposition 2.1.} {\it  Suupose that the marginal (frequency) probabilities
$p_a,..., p_{b^\prime}, p_{\bf a}, p_{\bf b}$ with respect to $\{ {\cal S}^{(M)}\}$
are well defined \footnote{Here, e.g.,
$p_{\bf a}(kl)= \lim_{M\to \infty} \nu( a=a_k, a^\prime= a^\prime_l;  {\cal S}^{(M)}).$}
and the probabilities
$p_{\bf b}$ with respect to $\{ {\cal S}_{\bf a}^{(M)}\}$ are well defined \footnote{Here
$p_{\bf b}(ij/kl)= \lim_{M\to \infty} \nu( b=b_i, b^\prime= b^\prime_j;  {\cal S}_{\bf a}^{(M)}),$
where ${\bf a} =(a_k, a_l^\prime).$} for each
value of the parameter ${\bf a}.$ Then there exists the limit (\ref{TRDL})
of the entanglement terms
$\delta^{(M)}\equiv \de(ij; {\cal S}^{(M)} , {\cal S}_{\bf a}^{(M)})$ and the (frequency)
probability $p_{\bf b}(ij)$ can be represented in the form (\ref{TRD}).}

\medskip

We are looking for a transformation of probabilities
which would give the possibility for representing the probabilities
\[p_{{\bf b}}(ij)={\bf P}_{\cal S}(b=b_i, b^\prime=b_j^\prime)\]
by using probabilities $p_{{\bf b/a}}(ij/kl).$ We underline that the latter
probabilities can be found experimentally.

We shall study the case, when the probability $p_{\bf b/a}$ can be factorized:
\begin{equation}
\label{F}
p_{\bf b/a}(ij/kl)=p_{b/a}(i;k)p_{b^\prime/a^\prime}(j;l) .
\end{equation}
Of course, the reader understand that (\ref{F}) is a kind of {\it independence condition.}
We remark that in the frequency framework (see R. von Mises [23])
independence  is not independence of events, but independence of experiments (independence
of collectives). The physical meaning of condition (\ref{F}) is  the standard one: independence of choices
of settings of measurement devices for
measurements over parts $\omega$ and $\omega^\prime$ of the composite system $w= (\omega,\omega^\prime).$

{\bf Remark 2.5.} (Locality and outcome independence) In principle, by analogy with Bell's
measure-theoretical probabilistic approach to locality we can interpret
the independence condition (\ref{F}) as a {\bf locality} condition. Of course,
the reader can be unsatisfied by such an approach to locality, since
space-variables are not at all involved into our considerations. But the
same critique can be directed against the original Bell's approach to locality:
he neither considered space-variables, see [24], [25] for detail. In principle we can call
our model {\bf frequency probabilistic local} model. On the other hand, it seems more natural
to speak just about {\it outcome independence} (as many authors do in the EPR-Bohm framework).

Under the outcome independence condition (\ref{F}) we have
\begin{equation}
\label{TRD1}
p_{\bf b}(ij)=\sum_{k,l=1}^2 p_{{\bf a}}(kl)
p_{b/a}(i/k) p_{b^\prime/a^\prime}(j/l)+ \bar{\de}_{{\bf b}/{\bf a}}(ij).
\end{equation}
We now consider one very special case, namely an ensemble ${\cal S}$ of{\it  anticorrelated systems.} Here:
\begin{equation}
\label{a}
p_{\bf a}(kk)={\bf P}(a=a_k, a^\prime=a_k^\prime)=0.
\end{equation}
In such a case the probability to obtain
the result $(a=a_1, a^\prime=a_1^\prime)$ or $(a=a_2, a^\prime=a_2^\prime)$ is
equal to zero (for example, we can consider values $a_1,a_1^\prime =+1$ and
$a_2,a_2^\prime= -1).$  In this case the entanglement term
$\delta(ij;{\cal S}, {\cal S}_{\bf a})$ contains only {\bf nondiagonal} nontrivial terms:
$$\de(ij;{\cal S}, {\cal S}_{\bf a}) \approx \nu_{{\bf a}}(12; {\cal S})
[\nu_{\bf b}(ij;{\cal S}_{0;{\bf a}}(12))-\nu_{\bf b}(ij;{\cal S}_{\bf a}(12))]
$$
$$
+ \nu_{\bf a}(21;{\cal S})[\nu_{\bf b}(ij;{\cal S}_{0;{\bf a}}(21))-
\nu_{\bf b}(ij;{\cal S}_{\bf a}(21))].
$$
We renormalize $\bar{\de}_{{\bf b/a}}(ij)=
\lim_{M\to \infty}\de^{(M)}(ij; {\cal S}, {\cal S}_{\bf a})$ by introducing a new entanglement
coefficient
$$
\lambda_{{\bf b/a}}(ij)=
\frac{\bar{\de}_{{\bf b/a}}(ij)}{2\sqrt{p_{\bf a}(12) p_{\bf a}(21)p_{b/a}(i/1)p_{b/a}(i/2)
p_{b^\prime/a^\prime}(j/1)p_{b^\prime/a^\prime}(j/2)}}\;.
$$
Thus we get the general probabilistic transformation for anti-correlated
systems (under the outcome independence condition (\ref{F})) :
$$
p_{\bf b}(ij)=p_{\bf a}(12)p_{b/a}(i/1)p_{b^\prime/a^\prime}(j/2)+
p_{\bf a}(21)p_{b/a}(i/2)p_{b^\prime/a^\prime}(j/1)
$$
\begin{equation}
\label{TRD2}
+
2\lambda_{{\bf b/a}}(ij) \sqrt{p_{\bf a}(12)p_{\bf a}(21)p_{b/a}(i/1)p_{b/a}(i/2)
p_{b^\prime/a^\prime}(j/1)p_{b^\prime/a^\prime}(j/2)}
\end{equation}
Entanglement coefficients $\lambda_{{\bf b/a}}(ij)$ can have various
magnitudes (depending on perturbation effects induced by
the transitions from ${\cal S}$ to ${\cal S}_{\bf a}(kl)).$  We consider various possibilities:

\medskip

1. {\bf The case of relatively small statistical perturbations.} Let all entanglement coefficients
$|\lambda_{{\bf b/a}}(ij)|\leq 1$.
We can represent these coefficients in the form $\lambda_{{\bf b/a}}(ij)=\cos \theta_{{\bf b/a}}(ij)$ where
$\theta_{{\bf b/a}}(ij) \in [0,2\pi)$
are some ``phases".\footnote{In our framework ``phases" $\theta_{{\bf b/a}}(ij)$ are purely probabilistic parameters.
This is just a new representation for the  entanglement coefficients $\lambda_{{\bf b/a}}(ij).$ Of course, as it often occurs
in probability theory, in some cases those probabilistic ``phases" could have a geometric meaning.}
Thus we get the following {\it trigonometric entanglement of probabilities} (compare to [18], [19] for noncomposite systems):
$$
p_{\bf b}(ij)=p_{\bf a}(12)p_{b/a}(i/1)p_{b^\prime/a^\prime}(j/2)+
p_{\bf a}(21)p_{b/a}(i/2)p_{b^\prime/a^\prime}(j/1)+
$$
\begin{equation}
\label{TRb}
2\cos \theta_{{\bf b/a}}(ij) \sqrt{p_{\bf a}(12)p_{\bf a}(21)p_{b/a}(i/1)p_{b/a}(i/2)p_{b^\prime/a^\prime}(j/1)p_{b^\prime/a^\prime}(j/2)}
\end{equation}
This is quantum-like case. In the conventional quantum formalism
this equation can be obtained by using a linear transformation in the tensor product
of two ${\bf C}$-linear spaces $H_1$ and $H_2.$

We now formulate the above result -- representation (\ref{TRb}) -- as the mathematical theorem:

\medskip

{\bf Theorem 1.} {\it Let conditions of proposition 1, the
outcome independence condition (\ref{F}), and the anticorrelation condition (\ref{a})
hold true. If all entanglement coefficients $\vert  \lambda_{{\bf b/a}}(ij)\vert \leq 1$
then the (frequency) probabilities
$p_{\bf b}(ij)$ can be represented in the form (\ref{TRb}).}

\medskip

By using the frequency probabilistic version of Bell's (measure-theoretical) terminology, see
Remark 2.5., we can formulate this result in the following way:

\medskip

{\bf Theorem 1a.} {\it In the local realists (frequency) framework
for anticorrelated systems with entanglement coefficients $\vert  \lambda_{{\bf b/a}}(ij)\vert \leq 1$
we have the representation
(\ref{TRb}) of the (frequency) probabilities for measurements on the parts $\omega$ and
$\omega^\prime$ of the  composite system $w=(\omega, \omega^\prime).$}

\medskip

2. {\bf Relatively large statistical perturbations.} Let all entanglement coefficients $|\lambda_{{\bf b/a}}(ij)|> 1.$
We can represent these coefficients in the form $\lambda_{{\bf b/a}}(ij)=\pm \cosh \theta_{{\bf b/a}}(ij),$
where $\theta_{{\bf b/a}}(ij)  \in (0, + \infty)$ are
``hyperbolic phases". Thus we get the following
{\it hyperbolic entanglement of probabilities}:
$$
p_{\bf b}(ij)=p_{\bf a}(12)p_{b/a}(i/1)p_{b^\prime/a^\prime}(j/2)+
p_{\bf a}(21)p_{b/a}(i/2)p_{b^\prime/a^\prime}(j/1)
$$
\begin{equation}
\label{TRCC}
\pm 2 \cosh \theta_{{\bf b/a}}(ij) \sqrt{p_{\bf a}(12)
p_{\bf a}(21)p_{b/a}(i/1)p_{b/a}(i/2)p_{b^\prime/a^\prime}(j/1)p_{b^\prime/a^\prime}(j/2)}
\end{equation}
This equation can be induced by a linear transformation in the tensor product of two
hyperbolic spaces $H_1$ and $H_2$  [18]  (modules over a two dimensional Clifford algebra).

\medskip

3. {\bf Mixed behaviour.} Let some $|\lambda_{{\bf b/a}}(ij)|\leq 1$ and some $|\lambda_{{\bf b/a}}(ij)| > 1.$
Here we get a mixture of trigonometric and hyperbolic entanglements. We do not know anything about
the possibility to represent such mixed probabilistic transformations in linear spaces (or modules).

\section{``Polarization probabilities"}
Let us consider in more detail the case of relatively small perturbation effects, namely
$\lambda_{{\bf b/a}}(ij)=\cos \theta_{{\bf b/a}}(ij).$ Let us consider an ensemble of anti-correlated with respect to
the ${\bf a}$-observable systems. We make the following simple remark:
\begin{equation}
\label{SS1}
p_{b/a}(1/1)+p_{b/a}(2/1)={\bf P}(b=b_1; S_a(1))+{\bf P}(b=b_2; S_a(1))=1 ,
\end{equation}
\begin{equation}
\label{SS2}
p_{b/a}(1/2)+p_{b/a}(2/2)={\bf P}(b=b_1; S_a(2))+{\bf P}(b=b_2; S_a(2))=1 .
\end{equation}
Here the probabilities $p_{b/a}(i/k)\equiv {\bf P}(b=b_i; S_a(k))$ are the
probabilities to find $b=b_i$ for an element $\omega \in S_a(k).$
Condition (\ref{SS1}), (\ref{SS2}) is well known condition of {\it stochasticity}
of the matrix of transition probabilities ${\bf P}(b/a)$. We remark that this condition is always
satisfied automatically. This is the conventional  condition of additivity of probability of alternatives for one fixed
context (in the mathematical formalism -- one fixed Kolmogorov probability space).
The same condition we have for $b^\prime$  and $a^\prime$. Thus we can always set:
$$
p_{b/a}(1/1)=\cos^2 \xi_1, p_{b/a}(2/1)(b/a)=\sin^2 \xi_1;
$$
$$
p_{b/a}(1/2)=\sin^2 \xi_2, p_{b/a}(2/2)=\cos^2 \xi_2
$$
with some (probabilistic) ``phases" $\xi_1, \xi_2 \in [0,\pi/2];$ we can also use a similar trigonometric
representation
for $b^\prime/a^\prime$ probabilities.

We study only the {\it symmetric case} in our further investigations:
\begin{equation}
\label{S}
p_{\bf a}(12)=p_{\bf a}(21)=1/2,
\end{equation}
so  ${\bf P}(a=a_1, a^\prime=a_2^\prime; {\cal S})={\bf P}(a=a_2, a^\prime=a_1^\prime)=1/2.$

In the symmetric case we have, for example,
that
$$
p_{{\bf b}}(11)= \frac{1}{2} [p_{b/a}(1/1)p_{b^\prime/a^\prime}(1/2)+
p_{b/a}(1/2) p_{b^\prime/a^\prime}(1/1)]
$$
$$
+ 2 \cos \theta_{11} \sqrt{p_{b/a}(1/1)p_{b^\prime/a^\prime}(1/2)p_{b/a}(1/2) p_{b^\prime/a^\prime}(1/1)} .
$$
Thus:
$$
p_{\bf b}(11)=\frac{1}{2} [\cos^2 \xi_1 \sin^2 \xi_2^\prime +
\sin^2 \xi_2 \cos^2 \xi_2^\prime]
$$
\begin{equation}
\label{PR}
+\cos \theta_{11} \cos \xi_1 \cos \xi_1^\prime \sin \xi_2 \sin \xi_2^\prime \;.
\end{equation}
This is the general expression for the trigonometric transformation of probabilities when the
matrixes ${\bf P}(b/a)=(p_{b/a}(i/j)), {\bf P}(b^\prime/a^\prime)=(p_{b^\prime/a^\prime}(i/j))$ are
{\it stochastic.}

We now consider more special case:
matrixes ${\bf P}(b/a)$ and ${\bf P}(b^\prime/a^\prime)$ are {\it double stochastic.}\footnote{We remark that
matrices in the EPR-Bohm experiment are double stochastic.}

Here, not only $p_{b/a}(1/j)+p_{b/a}(2/j)=1, j=1,2,$
but also $p_{b/a}(i/1) + p_{b/a}(i/2) =1, i=1,2.$
Thus we can set  $\alpha \equiv \xi_1= \xi_2$ and $\beta\equiv \xi_1^\prime=\xi_2^\prime.$

We get:
$$
p_{\bf b}(11)=\frac{1}{2}(\cos^2\al \sin^2 \beta + \sin^2 \al \cos^2\beta)+
\cos \theta_{11} \cos \al \cos \beta \sin \al \sin \beta
$$
$$
= \frac{1}{2} (\cos \al \sin \beta - \sin \al \cos \beta)^2+
(1+\cos \theta_{11}) \cos \al \cos \beta \sin \al \sin \beta
$$
$$
=
\frac{1}{2} \sin^2 (\al-\beta)+(1+\cos \theta_{11}) \cos \al \cos \beta \sin \al \sin \beta .
$$
In the same way we get that, for example,
$$
p_{{\bf b}}(12)=\frac{1}{2} [ p_{b/a}(1/1)p_{b^\prime/a^\prime}(2/2)+
p_{b/a}(1/2) p_{b^\prime/a^\prime}(2/1)]
$$
$$
+2 \cos \theta_{12}  \sqrt{p_{b/a}(1/1)p_{b^\prime/a^\prime}(2/2)p_{b/a}(1/2)p_{b^\prime/a^\prime}(2/1)}\;.
$$
Thus, for stochastic matrixes ${\bf P}_{b/a}$ and ${\bf P}_{b^\prime/a^\prime},$ we get:
$$
p_{\bf b}(12)=\frac{1}{2}(\cos^2 \xi_1 \cos^2 \xi_2^\prime + \sin^2 \xi_2 \sin^2 \xi_1^\prime)
$$
\begin{equation}
\label{EU}
+\cos \theta_{12} \cos \xi_1 \cos \xi_2^\prime \sin \xi_2 \sin \xi_1^\prime
\end{equation}
For double stochastic matrices ${\bf P}_{b/a}$ and ${\bf P}_{b^\prime/a^\prime}$, we get:
\begin{equation}
\label{EUY}
p_{\bf b}(12)=\frac{1}{2}(\cos^2\al \cos^2\beta + \sin^2 \al \sin^2 \beta)
+ \cos \theta_{12} \cos \al \cos \beta \sin \al \sin \beta
\end{equation}
$$
= \frac{1}{2} (\cos \al \cos \beta - \sin \al \sin \beta)^2 -
(1-\cos \theta_{12})\cos \al \cos \beta \sin \al \sin \beta,
$$
$$
...............
$$
We can formulate this result as the mathematical proposition:

{\bf Proposition 3.1.} {\it Let conditions of theorem 1 hold true, probabilities
are symmetric, (\ref{S}), and the matrices of transition probabilities are double
stochastic. Then we have the representations (\ref{EUY}),...,
of the (frequency) probabilities $p_{\bf b}(ij).$}

Suppose now that in experiments under consideration perturbation effects are such
that
$$
\cos\theta_{11}=-1 \; \mbox{and}\; \cos \theta_{12}=1.
$$
Here all entanglement coefficients $\lambda_{{\bf b/a}}(ij)$
have their maximal magnitudes:
\begin{equation}
\label{MG}
|\lambda_{{\bf b/a}}(ij)|=1.
\end{equation}
Thus {\it perturbations of probability distributions are very strong -- as strong as possible
in the case of trigonometric probabilistic behaviour.} In such a case we get, for
$\gamma = 2 \alpha$ and $\gamma^\prime = 2 \beta,$
\begin{equation}
\label{TP}
p_{\bf b}(ii)={\bf P}(b=b_i, b^\prime=b_i^\prime; {\cal S})=
\frac{1}{2}\sin^2 \frac{\gamma^\prime-\gamma}{2}
\end{equation}
\begin{equation}
\label{TP1}
 p_{\bf b}(ij)={\bf P}(b=b_i, b^\prime=b_j; {\cal S})=
\frac{1}{2}\cos^2 \frac{\gamma^\prime-\gamma}{2}, i \not = j.
\end{equation}

Finally, we formulate the following theorem:

{\bf Theorem 3.1.} {\it Let conditions of proposition 3.1 hold true and let
all entanglement coefficients have the maximal magnitude, $\vert \lambda_{{\bf b/a}}(ij)\vert=1.$
Then  the (frequency) probabilities $p_{\bf b}(ij)$ can be represented in
the form, (\ref{TP}), (\ref{TP1}), of the EPR-Bohm probabilities.}

\medskip

By using the frequency probabilistic version of Bell's (measure-theoretical) terminology, see
Remark 2.5., we can formulate this result in the following way:

\medskip

{\bf Theorem 3.1a.} {\it In the local realists (frequency) probabilistic framework
for anticorrelated systems with symmetric probability distributions, double stochastic
matrices of transition probabilities, and entanglement terms $\vert\lambda_{{\bf b/a}}(ij)\vert=1,$
the (frequency) probabilities for measurements on the parts $\omega$ and
$\omega^\prime$ of the  composite system $w=(\omega, \omega^\prime)$
can be represented in the form of the EPR-Bohm probabilities.}

Thus we have  obtained probabilities corresponding to
experiments of the EPR-Bohm type on polarization
measurements for correlated pairs of photons or spin measurements for electrons. To be closer to such experimental
situation, we can also assume, that $a, a^\prime, b, b^\prime=\pm 1.$

The condition of anti-correlation for the observable $a$  in this case is the following one:
\[{\bf P}(a=+1, a^\prime=+1;{\cal S})={\bf P}(a=-1, a^\prime=-1; {\cal S})=0 .\]
Thus we have
$$
a(\omega)a^\prime (\omega^\prime)=-1
$$
for almost all
pairs $w=(\omega, \omega^\prime)\in {\cal S}.$ This is precisely the situation that we have in
the EPR-Bohm experiments.

We now fix the direction $x$ and choose
the ${\bf a} =(a, a^\prime)$ measurement as the measurement of projections of
spins of correlated particles on the same axis $x.$ Here
$$
{\bf P}(a=+1,a^\prime=+1)={\bf P}(a =-1, a^\prime=-1) = 0 ,
$$
$$
{\bf P}(a =+1, a^\prime=-1)= {\bf P}(a=-1, a^\prime =+1) = \frac{1}{2}.
$$

We now choose in our general scheme $b=M_{\gamma}$ and $b^\prime=M_{\gamma^\prime}^\prime,$
where $M_{\gamma}, M_{\gamma^\prime}^\prime$ are
spin projections to directions having angles $\gamma, \gamma^\prime$,  with the $x$-direction.
In this case our general result (\ref{TP}), (\ref{TP1}) gives correct quantum mechanical
probabilities ${\bf P}(M_{\gamma}=\pm 1, M_{\gamma^\prime}^\prime=\pm 1).$

An important consequence
of our derivation is that EPR-Bohm probabilities might be in principle
obtained in experiments with (classical) {\it macroscopic systems.} We cannot present the
concrete experimental framework. But
in the contextual  (frequency) model there are no ``NO-GO"
theorems which would imply the impossibility of obtaining probabilities of
the EPR-Bohm form in experiments with classical systems.

We note that phases $\theta_{{\bf b/a}}(ij)$ are not independent. We have in the case of general stochastic transition matrixes:
$$
1=p_{{\bf b}}(11)+p_{{\bf b}}(22)+p_{{\bf b}}(12)+p_{{\bf b}}(21)
$$
$$
=\frac{1}{2}(\cos^2 \xi_1 \sin^2 \xi_2^\prime + \sin^2 \xi_2 \cos^2 \xi_1^\prime
+ \cos^2 \xi_2 \sin^2 \xi_1^\prime
$$
$$
+ \sin^2 \xi_1 \cos^2 \xi_2^\prime +
\cos^2 \xi_1 \cos^2 \xi_2^\prime + \sin^2 \xi_2 \sin^2 \xi_1^\prime
+\cos^2 \xi_2 \cos^2 \xi_1^\prime + \sin^2 \xi_1 \sin^2 \xi_2^\prime)
$$
$$
+\cos \theta_{11} \cos \xi_1 \cos \xi_1^\prime \sin \xi_2 \sin \xi_2^\prime + \ldots +
\cos \theta_{21} \cos \xi_2 \cos \xi_1^\prime \sin \xi_1 \sin \xi_2^\prime
$$
Thus we get
$$\cos \theta_{11} \cos \xi_1 \cos \xi_1^\prime \sin \xi_2 \sin \xi_2^\prime + \ldots +
+ \cos \theta_{21} \cos \xi_2 \cos \xi_2^\prime \sin \xi_1 \sin \xi_2^\prime =0 .
$$
In the case of double stochastic transition matrixes, we get:
$$
\cos \al \sin \al \cos \beta \sin \beta (\cos \theta_{11} + \cos \theta_{22} + \cos \theta_{12}+\cos \theta_{21})=0.
$$
If $\al, \beta \not = \frac{\pi}{2} k, k = 1, 2, \ldots,$ then
we get
$$
\cos \theta_{11} + \cos \theta_{12} + \cos\theta_{22} + \cos \theta_{21}=0.
$$

We recall that phases in the derivation of ``polarization probabilities" were the following ones:
$$
\cos \theta_{11}=-1, \cos \theta_{12}=1, \cos \theta_{22}=-1, \cos \theta_{21}=1 .
$$

\section{Contextuality, incompatibility, nonexistence of the simultaneous probability distribution}

Of course, our general statistical description of measurements over composite systems does not
provide a description of physical processes that could induce such probabilistic phases.
However, we demonstrated that only by taking into account  {\bf incompatibility}
of some physical observables for  composite systems we can derive
probabilities having the EPR-Bohm form in the local realist (frequency)
framework, compare to [22]. We remark that in our
probabilistic framework incompatibility
of physical observables is equivalent to {\bf contextuality of probabilities}, i.e.,
statistically nontrivial dependence of
probabilities on complexes of experimental physical conditions.

We have seen that to get the EPR-Bohm probabilities the coefficients of entanglement
(which give the measure of disturbance of probability distributions by measurements) should be
of the maximal magnitude.\footnote{It would be interesting to investigate the relation between
the magnitude of our frequency coefficient of entanglement and violation of Bell's inequality.}
 Here the crucial role is played by {\it incompatibility} of
observables ${\bf a}= (a, a^\prime)$ and ${\bf b}= (b, b^\prime)$ on composite systems.
By measuring of ${\bf a}$ (on a composite system)
we disturb very strongly the probability distribution of ${\bf b}.$ We emphasize again that we
do not speak about the influence of a measurement on one part of a composite system onto
another part of this system. There is discussed disturbance of a composite system.

For example, if we perform the measurements of the polarization in the $x$-direction
on both photons (in a EPR-Bohm pair), then by this act the probability distribution
of polarizations in other directions would be changes very strongly. In fact, as was remarked,
we need the coefficient of statistical entanglement of the maximal magnitude. Finally, we remark that
in our approach the coefficient of entanglement has merely the meaning of the coefficient of interference
between incompatible observables.

The Bell theorem
tells us that incompatible observables with so strong statistical disturbances cannot be realized
on a space with a single probability measure. However, in the frequency probabilistic approach
there are no reasons for the existence of such a measure for incompatible observables.\footnote{May be our
frequency probabilistic investigation can clarify the well known results of A. Fine and P. Rastal
[26].}

\newpage

{\bf Conclusion:} {\it  In fact, in this paper I did with Bell's approach
more or less the same thing as J. Bell did with approaches of von Neumann, Kohen and Specker,...
Both J. Bell and I speak about mathematical models of local realism. As J. Bell underlined
there can be proposed various mathematical models of realism (and local realism).
In particular, J. Bell denied approaches of von Neumann and  Kohen and Specker
and presented his own mathematical model. Many things which were
impossible in previous mathematical  models  became possible in Bell's model.
I use a mathematical model of realism which is essentially ``weaker'' than Bell's
model, see Remark 2.1. Therefore some things which were
forbidden in Bell's model are possible in my model. In particular, we can obtain
the EPR-Bohm probabilities in spite of Bell's theorem.}

I would like to thank L. Ballentine,
 S. Gudder,  W. De Muynck,
A. Holevo,    K. Gustafsson, I. Volovich
for fruitful (and rather critical) discussions.

{\bf References}

[1] P. A. M.  Dirac, {\it The Principles of Quantum Mechanics}
(Oxford Univ. Press, 1930).

[2] W. Heisenberg, {\it Physical principles of quantum theory.}
(Chicago Univ. Press, 1930).

[3]  N. Bohr, {\it Phys. Rev.,} {\bf 48}, 696-702 (1935).

[4] J. von Neumann, {\it Mathematical foundations
of quantum mechanics} (Princeton Univ. Press, Princeton, N.J., 1955).

[5] R. Feynman and A. Hibbs, {\it Quantum Mechanics and Path Integrals}
(McGraw-Hill, New-York, 1965).

[6] A. S. Holevo,  Probabilistic and statistical aspects of quantum theory.
(Nauka, Moscow, 1980; North Holland, 1982).

[7] A. Peres, {\em Quantum Theory: Concepts and Methods} (Kluwer Academic
Publishers, 1994).

[8]  G. Ludwig, {\it Foundations of quantum mechanics,} v.1. Springer-Verlag, Berlin
(1983).

[9] P. Busch, M. Grabowski, P. Lahti, {\it Operational Quantum Physics}
(Springer Verlag, 1995).

[10] E. Beltrametti  and G. Cassinelli, {\it The logic of Quantum mechanics.}
(Addison-Wesley, Reading, Mass., 1981).

[11] Holevo, A.S.. Statistical structure of quantum theory.
(Lect. Notes Phys., 67, Berlin, 2001).

[12] A.Yu. Khrennikov, {\it Interpretations of
probability} (VSP Int. Publ., Utrecht, 1999).

[13] R. P. Feynman, R. B. Leighton and M. Sands, {\it The Feynman
lectures on physics.} {\bf 3}, Addison-Wesley, Reading, Massachusets (1965).

[14]  A. Zeilinger, On the interpretation and philosophical foundations of
quantum mechanics.
in {\it  Vastakohtien todellisuus.} Festschrift for K.V. Laurikainen.
U. Ketvel et al. (eds). (Helsinki Univ. Press, 1996).

[15] J. S. Bell, {\it Speakable and unspeakable in quantum mechanics.}
(Cambridge Univ. Press, 1987).

[16] N. Bohr, {\it Niels Bohr: Collected works.} {\bf 1-10}
(Elsevier, Amsterdam, 1972-1996).

[17] A. Plotnitsky, {\it Quantum atomicity and quantum information: Bohr, Heisenber,
and quantum mechanics as an information theory.} Proc. Int. Conf. "Quantum Theory: Reconsideration
of Foundations". Ser. Math. Modelling in Phys., Engin., and Cogn. Sc., ed.: A. Khrennikov, 309-342,
V\"axj\"o Univ. Press, 2002.

[18] A. Yu. Khrennikov,  {\it J. Phys.A: Math. Gen.,} {\bf 34}, 9965-9981 (2001);
 {\it Il Nuovo Cimento,} {\bf B 117,}  267-281  (2002);
{\it J. Math. Phys.}, {\bf 43}, 789-802 (2002).

[19]  L. Accardi, The probabilistic roots of the quantum mechanical paradoxes.
{\em The wave--particle dualism.  A tribute to Louis de Broglie on his 90th
Birthday,} ed. S. Diner, D. Fargue, G. Lochak and F. Selleri
(D. Reidel Publ. Company, Dordrecht, 297--330, 1984);

L. Accardi, {\it Urne e Camaleoni: Dialogo sulla realta,
le leggi del caso e la teoria quantistica.} (Il Saggiatore, Rome, 1997).

L. Accardi and M. Regoli,  Locality and Bell's inequality.
{\it Foundations of Probability and Physics,}
{\it Q. Prob. White Noise Anal.}, {\bf 13}, 1-28 (WSP, Singapore, 2001).

[20] S. P. Gudder, Trans. AMS 119, 428 (1965);  {\it J. Math Phys.,} {\bf 25}, 2397- 2401 (1984).

S. P. Gudder, {\it Axiomatic quantum mechanics and generalized probability theory}
(Academic Press, New York, 1970).

S. P. Gudder, An approach to quantum probability,
in:  A. Yu. Khrennikov (Ed.),
{\it Foundations of Probability and Physics, Q. Prob. White Noise Anal.},  {\bf 13,}
147-156, WSP, Singapore, 2001.

[21] J.F. Clauser , M.A. Horne, A. Shimony, R. A. Holt,
Phys. Rev. Letters, {\bf 49}, 1804-1806 (1969);
J.F. Clauser ,  A. Shimony,  Rep. Progr.Phys.,
{\bf 41} 1881-1901 (1978).
 D. Home,  F. Selleri, Nuovo Cim. Rivista, {\bf 14},
2--176 (1991). H. P. Stapp, Phys. Rev., D, {\bf 3}, 1303-1320 (1971);
P.H. Eberhard, Il Nuovo Cimento, B, {\bf 38}, N.1, 75-80(1977); Phys. Rev. Letters,
{\bf 49}, 1474-1477 (1982);
A. Peres,  Am. J. of Physics, {\bf 46}, 745-750 (1978).
P. H. Eberhard,  Il Nuovo Cimento, B,
{\bf 46}, N.2, 392-419 (1978); J. Jarrett, Noûs, {\bf 18},
569 (1984).

[22] W. De Baere,  {\it Lett. Nuovo Cimento,} {\bf 39}, 234 (1984);
{\bf 40}, 448 (1984).

W. De Muynck and W. De Baere W.,
Ann. Israel Phys. Soc., {\bf 12}, 1-22 (1996);

W. De Muynck, W. De Baere, H. Marten,
Found. of Physics, {\bf 24}, 1589--1663 (1994);
W. De Muynck, J.T. Stekelenborg,  Annalen der Physik, {\bf 45},
N.7, 222-234 (1988).

[23] R.  von Mises, {\it The mathematical theory of probability and
 statistics} (Academic, London, 1964).

[24] A. Khrennikov, I. Volovich,  A. Yu. Khrennikov, I. Volovich, Local Realism, Contextualism  and
 Loopholes in Bell`s Experiments. quant-ph/0212127.

[25] A.Yu. Khrennikov, I.V. Volovich,   {\it
Quantum Nonlocality, EPR Model, and Bell`s Theorem}, Proceedings of the
3nd Sakharov conference on physics, Moscow, 2002, World Sci., vol.2, pp.269-276 (2003).

[26] A. Fine,  Phys. Rev. Letters, {\bf 48}, 291--295 (1982);
P. Rastal, Found. Phys., {\bf 13}, 555 (1983).

\end{document}